\documentclass[prl,twocolumn,showkeys,nofootinbib,showpacs,superscriptaddress]{revtex4-1}

\usepackage{graphicx}
\usepackage{url}
\usepackage{amsmath}
\usepackage{amssymb}
\usepackage{bm}
\usepackage{dsfont}
\usepackage{graphicx}
\usepackage{subfigure}
\usepackage{dcolumn}

\newcommand{\ri}{\mathrm{i}}
\newcommand{\re}{\mathrm{e}}
\newcommand{\rd}{\mathrm{d}}
\renewcommand{\ao}{\hat{a}^{\phantom\dag}}
\renewcommand{\aa}{\hat{a}^{\dag}}
\newcommand{\no}{\hat{n}}
\newcommand{\Ho}{\hat{H}}

\newcommand{\la}{\langle}
\newcommand{\ra}{\rangle}
\newcommand{\bA}{{\bm A}}
\newcommand{\br}{{\bm r}}

\newcommand{\bF}{{\bm F}}

\newcommand{\bx}{{\bm x}}

\begin{document}

\title{Tunable gauge potential for neutral and spinless particles in driven lattices}
\author{J. Struck}
\affiliation{Institut f\"ur Laserphysik, Universit\"at Hamburg, D-22761 Hamburg, Germany.}
\author{C. \"Olschl\"ager}
\affiliation{Institut f\"ur Laserphysik, Universit\"at Hamburg, D-22761 Hamburg, Germany.}
\author{M. Weinberg}
\affiliation{Institut f\"ur Laserphysik, Universit\"at Hamburg, D-22761 Hamburg, Germany.}
\author{P. Hauke}
\affiliation{Institut de Ci\`encies Fot\`oniques, Mediterranean Technology Park, Av. Carl Friedrich Gauss 3, E-08860 Castelldefels, Barcelona, Spain.}
\author{J. Simonet}
\affiliation{Institut f\"ur Laserphysik, Universit\"at Hamburg, D-22761 Hamburg, Germany.}
\author{A. Eckardt}
\affiliation{Max-Planck-Institut f\"ur Physik komplexer Systeme, N\"othnitzer Str. 38, D-01187 Dresden, Germany.}
\author{M. Lewenstein}
\affiliation{Institut de Ci\`encies Fot\`oniques, Mediterranean Technology Park, Av. Carl Friedrich Gauss 3, E-08860 Castelldefels, Barcelona, Spain.}
\affiliation{ICREA-Instituci\`o Catalana de Recerca i Estudis Avan\c{c}ats, Lluis Companys 23, E-08010 Barcelona, Spain.}
\author{K. Sengstock}
\email{klaus.sengstock@physnet.uni-hamburg.de}
\affiliation{Institut f\"ur Laserphysik, Universit\"at Hamburg, D-22761 Hamburg, Germany.}
\author{P. Windpassinger}
\affiliation{Institut f\"ur Laserphysik, Universit\"at Hamburg, D-22761 Hamburg, Germany.}

\begin{abstract}
We present a universal method to create a tunable, artificial vector gauge potential for neutral particles trapped in an optical lattice. The necessary Peierls phase of the hopping parameters between neighboring lattice sites is generated by applying a suitable periodic inertial force such that the method does not rely on any internal structure of the particles. We experimentally demonstrate the realization of such artificial potentials, which generate ground state superfluids at arbitrary non-zero quasi-momentum.
We furthermore investigate possible implementations of this scheme to create tuneable magnetic fluxes, going towards model systems for strong-field physics.
\end{abstract}

\maketitle

\begin{figure}
\centering
\includegraphics[width=8.5cm]{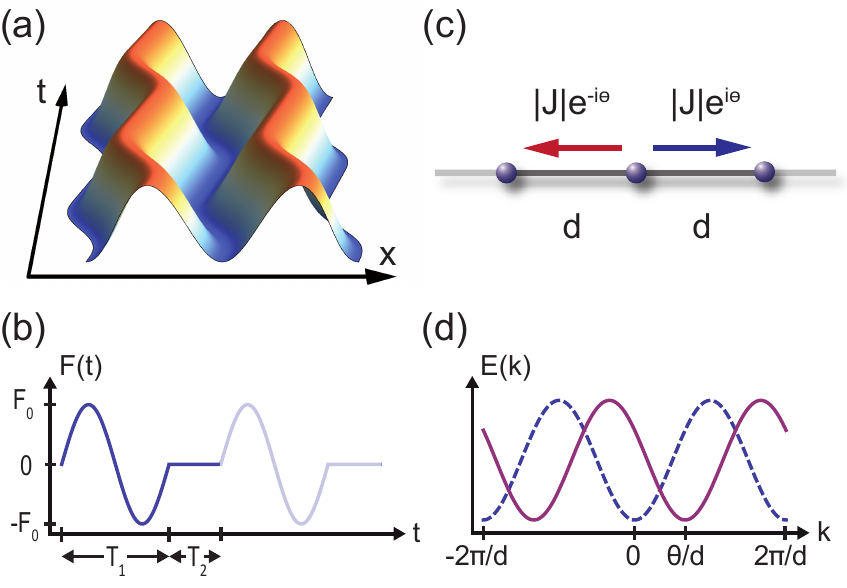}
\caption{\label{fig1} Periodic driving of the 1D lattice. \textbf{(a)} Time-periodic movement of the lattice in real space. \textbf{(b)} Resulting periodic inertial force of zero mean. $T_1/T_2$ is the asymmetry parameter. \textbf{(c)} Realized complex tunneling elements on a 1D lattice, with spacing $d$. \textbf{(d)} Effective single-particle dispersion relation. A complex phase induces a shift of $E_{\mathrm{eff}}(k)$ towards $k_{\mathrm{min}}$\,=\,$\theta/d$ (solid line).}
\end{figure}

First introduced in electromagnetism, gauge fields play a central role in the description of interactions in physics, from particle physics to condensed matter. Currently there is a large interest to introduce gauge fields into model systems in order to study fundamental aspects of physics \cite{Lewenstein2007}. Especially the emulation of synthetic electric and magnetic fields for of ultracold atomic systems is crucial in order to extend their proven quantum simulation abilities further, e.g. to quantum Hall physics or topological insulators. In this context, the analogy between inertial and Lorentz forces triggered the simulation of homogeneous artificial magnetic fields using rapidly rotating trapped ultracold gases \cite{Dalibard2004}. Recently, several proposals (\cite{Zoller2003, DalibardRMP2011} and references therein) and experimental realizations focused on the simulation of a gauge vector potential $\bm{A}$ either in a bulk system \cite{SpielmanPRL2009,Spielman2009} or in optical lattices \cite{Bloch2011, Spielman2012}.
The realized schemes exploit the Berry phase which arises when the atomic ground state is split in several space-dependent sublevels, as in the presence of an electromagnetic field. Hence they rely on the coupling between internal and external degrees of freedom induced by laser fields.
\newline
Here we demonstrate the generation of artificial gauge potentials for neutral atoms in an optical lattice without any requirements on the specific internal structure. As the realized scheme only relies on the trapability of the particle, it can be very widely applied to many atomic systems, in principle also to molecules and other complex particles. It is particularly interesting for fermionic systems, where in a many-body state governed by Pauli principle, the use of internal degrees of freedom often lead to conflicts with the creation of gauge potentials. As an important additional benefit, the internal degrees of freedom of the particles can be addressed independently, e.g. by real magnetic fields or microwave excitations.
\newline
In general, the presence of a gauge vector potential modifies the kinetic part of the Hamiltonian describing the system. In a lattice, an artificial field can then be simulated by engineering a complex tunneling parameter $J=|J|\cdot e^{i \theta}$, where $\theta$ is the Peierls phase.
\newline
The central approach here is to control this phase via a suitable forcing of the lattice potential, acting at the single-particle level.
We describe the general scheme for the creation of an artificial gauge potential and experimentally demonstrate the realization of a tunable vector gauge potential in 1D. We analyze the dynamical processes leading to the relaxation towards a superfluid ground-state at any desired, finite quasi-momentum. In a 2D lattice the described forcing can be used to create artificial magnetic fluxes smoothly tunable between $-\pi$ and $\pi$. This opens the possibility to emulate strong-field physics with cold atom systems, in a regime currently not accessible in condensed matter because of experimental limitations on the magnetic field strength.
\newline
For the general scheme, we consider a system of ultracold atoms in a deep optical lattice described by the Hubbard-type tight-binding Hamiltonian:
\begin{equation}
\Ho(t)= -\sum_{\la ij\ra} J_{ij}\aa_i\ao_j +\sum_i v_i(t)\no_i +\Ho_\text{on-site}.
\end{equation}
Here $\aa_i$, $\ao_i$ and $\no_i$ denote the creation, annihilation and number operator for a particle (boson or fermion) of mass $m$ localized at site $i$ at $\br_i$; $J_{ij}>0$ quantifies the tunneling between neighboring sites, and $\Ho_\text{on-site}$ comprises time-independent on-site terms describing, e.g. interactions or a trapping potential. The system is driven by fast, off-resonant time-periodic potential modulations $v_i(t)=v_i(t+T)$ of zero time average $\la v_i\ra_T\equiv\frac{1}{T}\int_0^T\!\rd t\,v_i(t)$ with period $T$. The case of a homogeneous inertial force $\bF=-m\ddot{\bx}$, which is created by shaking the lattice along the periodic orbit $\bx(t)$ in space (see Fig. \ref{fig1}(a)), is described by $v_i=-\br_i\cdot\bF$. Under the conditions described in Ref.\ \cite{Eckardt2010}, one can show that the driven system is to good approximation described by the effective time-independent Hamiltonian:
\begin{equation}
\label{eq:Heff}
\Ho_\text{eff} = -\sum_{ij}|J^\text{eff}_{ij}|\re^{\ri\theta_{ij}}\aa_i\ao_j+\Ho_\text{on-site}
\end{equation}
The important terms here are the complex tunneling parameters $|J^\text{eff}_{ij}|\re^{\ri\theta_{ij}}=J_{ij}\la\re^{\ri(\chi_j-\chi_i)/\hbar}\ra_T$ (see Fig. \ref{fig1}(c)), where $\chi_i(t)=-\int_{t_0}^t\!\rd t'\, v_i(t') + \la\int_{t_0}^t \!\rd t'\, v_i(t')\ra_T$.
For sinusoidal forcing, such a dynamic modification of tunneling is restricted to $\theta_{ij}=0$ or $\pi$ \cite{Dunlap1986,Hanggi1992,Holthaus1992,Eckardt2005}. It has recently been observed in several experiments \cite{Arimondo2007,Oberthaler2008,Arimondo2009,Struck2011}. We will now show that by suitable driving, the Peierls phases $\theta_{ij}$ can in fact be smoothly tuned to any value ($\theta \in (0,2\pi]$). This allows one to engineer vector potentials that can give rise to both artificial electric and magnetic fields. Similar ideas have recently been presented for a Bose-Fermi mixture in a triangular lattice \cite{Sacha2011}.
\newline
The Peierls phase $\theta_{ij}$ can be varied smoothly, even though $\la\chi_i\ra_T=0$. Indeed $\la\re^{\ri(\chi_j-\chi_i)}\ra_T$ can be complex provided the forcing breaks two symmetries that are also known to prevent ratchet-type rectification in classical \cite{Flach2000} and quantum \cite{Hanggi2007} lattice systems, namely: (a) reflection symmetry for a suitable time $\tau$, i.e. $v_i(t-\tau)=v_i(-t-\tau)$  and (b) shift (anti)symmetry, i.e. $v_i(t-T/2)=-v_i(t)$. This requires forcing with more than one frequency. The forcing function used in our experiment (see Fig. \ref{fig1}(b)) induces a monotoneously increasing Peierls phase
$\theta_{ij}=\arg(\la\re^{\ri m\dot\bx\cdot(\br_j-\br_i)}\ra_T)$ with the amplitude of the force.
\newline
The Peierls substitution directly links this phase to a gauge vector potential $\bm{A}$ via $\theta_{ij}=\int_{\br_j}^{\br_i}\!\rd\br\cdot\bA/\hbar$ (integrated along a straight path), which allows to create artificial electric or magnetic fields. The modification of the Peierls phase in time results in an artificial electric force $\bF_E=-\dot\bA$. It will be shown at the end of this manuscript that our scheme allows the emulation of a vector potential that gives rise to a finite artificial magnetic flux through the elementary plaquettes of 2D lattices.
\newline
\begin{figure}[b]
\centering
\includegraphics[width=8.5cm]{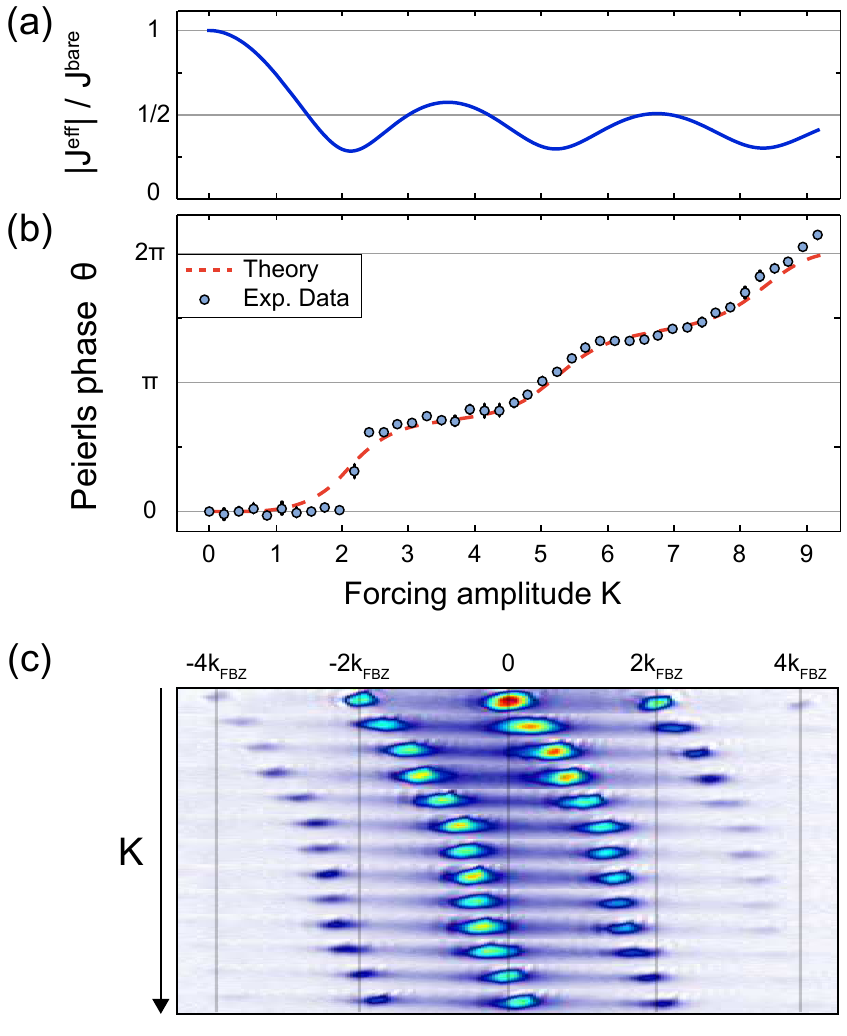}
\caption{\label{fig2} Creation of complex tunneling matrix elements. \textbf{(a)} Absolute value of the tunneling parameter obtained from equation (\ref{eq:ComplexJ}) for our experimental parameters ($T_1+T_2$=1\,ms and $T_1/T_2$=2.1). \textbf{(b)} The measured Peierls phases in a 1D driven optical lattice for different values of the forcing amplitude $K$ are depicted as circles. The dashed red curve corresponds to the theoretically expected values (equation (\ref{eq:ComplexJ})). \textbf{(c)} Quasi-momentum distribution of the BEC after 27\,ms time-of-flight for different values of $K$. The Peierls phase as a function of $K$ is deduced from the observed shifts of the interference patterns.}
\end{figure}
To demonstrate the power of this scheme, we have emulated a vector potential for a Rubidium Bose-Einstein condensate (BEC) in a one-dimensional lattice. The trapped particles are accelerated via frequency modulation of one of the lattice beams, a technique which is experimentally straightforward. The resulting inertial force is comprised of a train of sinusoidal pulses separated by periods of rest of periodicity $T$=$T_1+T_2$=\,1\,ms as depicted in Fig. \ref{fig1}(b). The zero mean value of this force prevents the transfer of a net-acceleration to the lattice. This leads to a renormalized tunneling matrix element of the form:
\begin{equation}
\label{eq:ComplexJ}
\frac{J^{\mathrm{eff}}(K)}{J^{\mathrm{bare}}}= \frac{T_2}{T} e^{i K \frac{T_1}{T}} + \frac{T_1}{T} J_0(K) e^{-iK \frac{T_2}{T}}
\end{equation}
where $T_1/T_2$ is the asymmetry parameter; $K$ is the forcing amplitude (see Appendix) and $J_0(K)$ the zero-order Bessel function. From equation (\ref{eq:ComplexJ}), the induced Peierls phase $\theta$ in the effective tunneling element can be calculated. For $T_2/T \rightarrow 0$, one recovers the harmonic driving introduced in \cite{Eckardt2005} which only allows to control magnitude and sign of $J^\mathrm{eff}$. In the opposite limit of small $T_1/T$, the tunneling amplitude $|J^{\mathrm{eff}}|$ is only marginally affected whereas the phase $\theta$ depends linearly on the forcing amplitude $K$. In order to avoid unwanted excitations of the system induced by strong forcing, we have chosen an intermediate value  $T_1/T_2=2.1$ for the experiment realized here. Note however that for such an asymmetry parameter the Peierls phase depends non-linearly on $K$ (see Fig. \ref{fig2}(b)).
\newline
The thereby introduced gauge potential $A=\hbar\theta/d$ manifests itself via a shift of the dispersion relation by $\theta/d$ as depicted in Fig. \ref{fig1}(d). This follows directly from the effective Hamiltonian (\ref{eq:Heff}), whose eigenstates are Bloch waves with the dispersion relation $E(k)=-2 |J^{\mathrm{eff}}|\cos(kd-\theta)$. The gauge potential therefore allows for the generation of superfluid ground-states (with group velocity $v_g=dE/dk=0$) at finite and tunable quasi-momentum $k=A/\hbar$. As will be detailed later on, we experimentally observe the relaxation of the condensate quasi-momentum towards the minimum of the effective dispersion relation. Therefore, the imprinted Peierls phase can be directly read out from the quasi-momentum distribution revealed in time-of-flight after a sudden switch off of the lattice and the external potential.
\newline
As a central result Fig. \ref{fig2}(b) shows the experimental data together with the theoretical predictions from equation (\ref{eq:ComplexJ}). After increasing the forcing amplitude slowly (in up to 120\,ms) to the desired value, the corresponding quasi-momentum distribution was recorded. From the obtained time-of-flight images, examples of which are shown in Fig. \ref{fig2}(c), we extract the Peierls phases $\theta$ (see Appendix). We observe an excellent agreement between experiment and theory, thus proving the controlled generation of an arbitrary vector gauge potential encoded into the Peierls phase $\theta \in [0,2\pi]$. In addition, the experimental images demonstrate the large degree of coherence maintained in the atomic sample throughout the shaking process. As an additional feature, Fig. \ref{fig2}(a) shows that the Peierls phase allows now to invert the sign of the effective tunneling element without crossing $|J^\mathrm{eff}|$=\,0 via the rotation in the complex plane.
\newline
In the following we will discuss the details of the relaxation of the system towards non-zero quasi-momenta superfluid states, allowing for the described direct measurement of the Peierls phase. Note that for an homogeneous and non-interacting system, the initial Bloch wave at $k_i$\,=\,0 remains an eigenstate of the effective Hamiltonian. Thus no transfer to states with $k$\,$\neq$\,0 is expected after the shaking is turned on. However, since we are working with interacting bosons and an external harmonic confinement more effects come into play.
\newline
\begin{figure}[t]
\centering
\includegraphics[width=8.5cm]{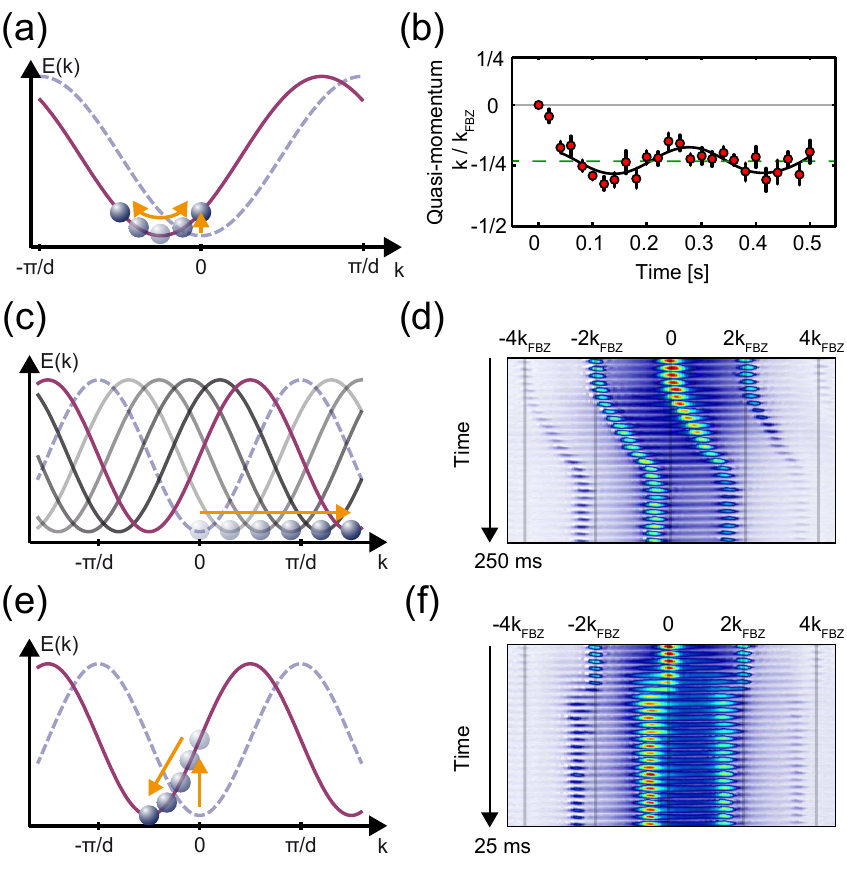}
\caption{\label{fig3} Time-resolved measurements. \textbf{(a),(b)} A quench to a Peierls phase of -$\pi$/4 within 1\,ms induces an excitation of the dipole mode, leading to oscillations of the quasimomentum as depicted in \textbf{(a)} and observed in the time-resolved measurement of the quasi-momentum up to 0.5\,s.\textbf{(b)}. The plain curve is a fit to a sinusoid from which one obtains $f_\mathrm{exp}$\,=\,3.6\,$\pm$\,0.4\,Hz, in perfect agreement with the expected dipole frequency of the dressed condensate. The final value of the Peierls phase is depicted as a dashed line. \textbf{(c)-(f)} Shift of the quasi-momentum towards a final Peierls phase of 3$\pi$/2 for a slow increase of the forcing amplitude over 160\,ms \textbf{(c),(d)} or a sudden quench within 1\,ms \textbf{(e),(f)}. The left graphs depict the center of mass dynamics in $k$-space in both cases. On the right, series of time-of-flight images are reported as a function of time. The time-steps are 10\,ms and 1\,ms, respectively.}
\end{figure}
When the gauge potential is ramped up from 0 to $A_f$, the condensate acquires a non-zero group velocity, reflecting the presence of an artificial electric force $F_E$\,=\,-$\dot{A}$. This velocity induces a displacement of the condensate's center of mass position $x_c$ in the harmonic potential of frequency $f$ (see Appendix). The resulting restoring force induces oscillations both in position and momentum-space (see Fig. \ref{fig3}(a)). In Fig. \ref{fig3}(b), we report a time-resolved measurement of the condensate quasi-momentum after a quench to a final Peierls phase of -$\pi$/4. The oscillations around the final quasi-momentum result from an excitation of the dipole mode: the measured frequency of 3.6\,$\pm$\,0.4\,Hz perfectly matches the expected dressed condensate frequency $\sqrt{m/m^*} f$ for particles having an effective mass $m^*$ in the lattice of 10\,$\pm$1\,$E_{\mathrm{rec}}$ depth with a tunneling amplitude of 0.3 $J_{\mathrm{bare}}$ ($f_\mathrm{theo}$=\,3.5\,$\pm$\,0.5\,Hz). The coupling to non-zero quasi-momenta results thus from the underlying harmonic trapping potential.
\newline
In addition, this center of mass dynamics is subjected to several damping mechanisms induced by the trap anharmonicity or the lattice discreteness, which leads to a coupling to other collective modes and therefore to the relaxation of the BEC towards the new equilibrium state. Therefore the duration of the ramp from 0 to $A_f$ has to be compared with the timescale of those relaxation mechanisms.
In Fig. \ref{fig3} we compare time-resolved measurements of the quasi-momentum distribution for a slow ramp (Fig. \ref{fig3}(d)) of $A$ to a final Peierls phase $\theta$=3$\pi$/2, with a sudden quench (Fig. \ref{fig3}(f)). As the gauge field is slowly increased, the BEC follows the shift of the dispersion relation minimum, as depicted in Fig. \ref{fig3}(c). For the quench, on the contrary, for which the shift of the dispersion relation occurs within one ms, the system cannot follow and thus relaxes into the nearest minimum of the effective band structure (see Fig \ref{fig3}(e)). For the chosen value this minimum lies on the left with respect to the original $k$=0 peaks and we thus find the BEC at $k$\,=\,-$\pi$/2$d$. This demonstrates clearly that in presence of these relaxation mechanisms, the forcing does not induce a net particle current in the lattice unlike for ratchets but allows the engineering of ground-state superfluids at arbitrary non-zero quasi-momenta.
\newline
Our scheme to generate complex tunneling offers fascinating possibilities to emulate gauge fields in higher dimensions. In the following, few examples based on lattice shaking and modulated superlattices will be described. An artificial magnetic field is characterized by a finite magnetic flux through an elementary plaquette $P$ of the lattice, given by the sum $\Phi_P=\theta_{ij}+\theta_{jk}+\cdots+\theta_{li}\mod 2\pi\in(-\pi,\pi]$ taken around $P$. Homogeneous forcing, implemented by lattice shaking, cannot give rise to a magnetic flux through plaquettes having pairwise parallel edges like square or hexagonal. It can, however, lead to a non-zero flux $\phi_\Delta$ through a triangular plaquette; $\phi_\Delta=\theta(|\bF_0|d)-2\theta(|\bF_0|d/2)$ for $\bF_0$ the amplitude of the force oriented parallel to one edge. As depicted in Fig. \ref{fig4}(a), continuously tunable staggered fluxes are realized ($\phi_\nabla=-\phi_\Delta$ for the inverted plaquette).
\newline
\begin{figure}
\centering
\includegraphics[width=8.5cm]{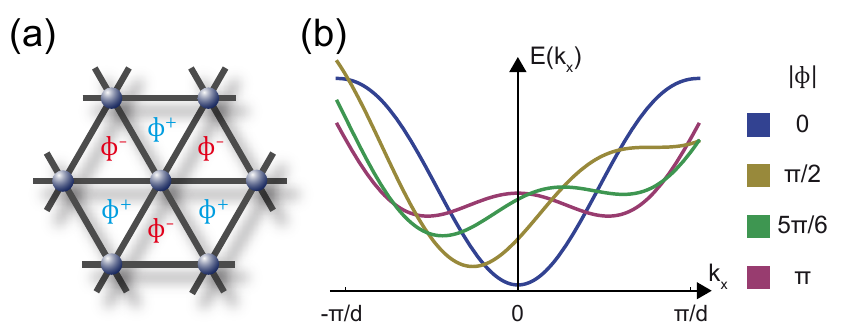}
\caption{\label{fig4} Creation of magnetic fluxes in triangular lattices. \textbf{(a)} Staggered magnetic fluxes obtained in a 2D triangular lattice. \textbf{(b)} Dispersion relation for different values of the magnetic flux $|\Phi|$ through a plaquette.}
\end{figure}
Our method enables one to continuously control the degree of frustration from zero ($\phi_\Delta=0$) to maximum ($\phi_\Delta=\pi$) in lattice geometries like triangular or Kagom\'e \cite{Becker2010,StamperKurn2012}. This control can be used to adiabatically prepare and study exotic quantum phases that are expected to appear in the strongly frustrated limit. Increasing the flux in a triangular lattice, frustration is reflected in the fact that the dispersion relation develops two minima that eventually, for $\pi$-flux, become degenerate (see Fig. \ref{fig4}(b)).
In a triangular lattice with $\pi$-flux spontaneous symmetry-breaking has recently been observed in an experimental simulation of classical magnetism \cite{Struck2011}. With the presented scheme, one is now able to study the influence of a small symmetry-breaking perturbation and to observe the many-body dynamics initiated by preparing a system in the upper minimum.
\newline
Tunable magnetic fluxes through e.g. square or hexagonal plaquettes can also be obtained via inhomogeneous forcing. A simple way of doing this, is to create an oscillating super-lattice as will be described elsewhere.
\newline
Beyond the engineering of the hopping element, another essential feature of our scheme is the possibility to address the atomic internal degrees of freedom independently. Here a spin-dependent hexagonal lattice \cite{Soltan2011} constitutes a promising system. In such a geometry the interplay of next-nearest neighbor coupling, complex tunneling matrix elements and spin are particularly interesting in connection with Haldane's model for instance \cite{Ripoll2011,Wang2008}.
\newline
The very general method presented in this manuscript should allow for large sets of new gauge fields schemes for various particle classes in optical lattices.
\newline
We thank the Deutsche Forschungsgemeinschaft (FOR 801, GRK 1355, SFB 925) and the Landesexzellenzinitiative Hamburg, which is supported by the Joachim Herz Stiftung, for funding. P.H. and M.L. are grateful for support through the European Research Council grant QUAGATUA, AAII-Hubbard and Caixa Manresa.

\section{Appendix}

\subsubsection{1D OPTICAL LATTICE:}
The 1D optical lattice is generated by two independent, phase stabilized laser beams at $\lambda\,$=\,830\,nm, intersecting at an angle of 117$^\circ$ in the $xy$ plane, which results in a lattice spacing of $d$\,=\,486\,nm along $z$. The radial dimensions are only weakly confined by an optical dipole trap of trap frequency $f_z$\,=\,18\,$\pm$\,2\,Hz and $f_x$\,=\,45-55\,Hz. We load a Bose Einstein condensate of Rubidium 87 atoms into the lattice, which results in an occupation of around 80 pancake-like lattice sites. The lattice depth is 10\,$\pm$1\,$E_\mathrm{rec}$, where $E_\mathrm{rec}$\,=\,2$\hbar^2\pi^2/\lambda^2m$\,=\,3.33\,kHz. This results in a bare tunneling rate of roughly 0.007\,$E_\mathrm{rec}$. The BEC is in the weakly interacting regime where $U/J<<$1, with on-site interaction $U$.
\newline

\subsubsection{LATTICE SHAKING:} The lattice is accelerated along the $z$ axis by modulating the frequency of one of the lattice beams with an acousto-optical modulator. We shake the lattice time-periodically with $\tilde{\omega}/2 \pi$ =\,1\,kHz, as the modulation frequency has to be fast with rsespect to the atomic motion but small compared to band excitation energies.
\newline
The inertial force acting on the atoms in the lattice frame consists of a train of sinusoidal pulses and is defined over a period as:
\begin{eqnarray}
\label{eq:F}
F(t)=\begin{cases}

 F_0 \sin{(\omega_1 t)} & \text{for} ~ 0<t<T_1=\frac{2\pi}{\omega_1}\\
  0 & \text{for} ~ T_1<t<T=\frac{2\pi}{\tilde{\omega}}
\end{cases}
\end{eqnarray}

The forcing parameter $K$ is defined as:
\begin{equation}
K=\frac{F_0 d}{\hbar \omega_1}= \frac{m d^2 \nu}{\hbar}
\end{equation}
where $\nu$ is the amplitude of the frequency modulation, ranging here from 0 to 30\,kHz.
\newline
Many different analytical functions may be chosen for the time-dependent inertial force.

\begin{figure}[h]
\centering
\includegraphics[width=8cm]{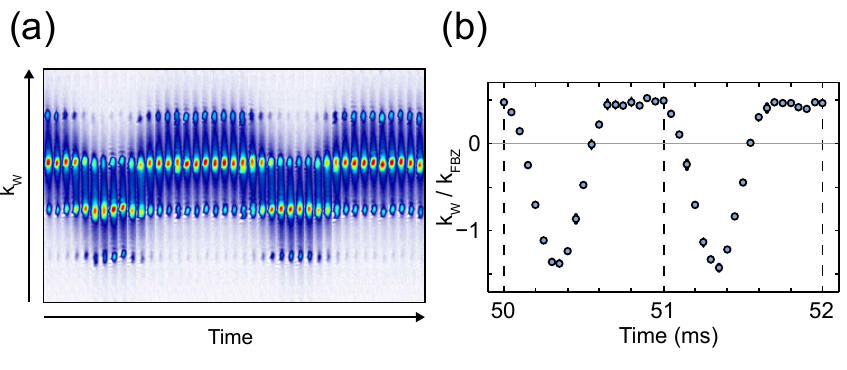}
\caption{\label{fig_SM} \textbf{(a)} Absorption images taken after 27\,ms time-of-flight over two periods of forcing. The center-of-mass oscillates under the effect of the forcing. The quasi-momentum, on the contrary, remains at any time described by the effective Hamiltonian. \textbf{(b)} Plot of the fitted momentum of the background (resembling in good approximation the Wannier envelope), denoted $k_W$, as a function of time.}
\end{figure}

In order to induce complex tunneling parameters, the function should break the relevant symmetries (a) and (b) described in the text. For this, the frequency spectrum of the force has to contain more than one component.
\newline
A major advantage of the function defined in (\ref{eq:F}) is the control of the sharpness of the Peierls phase dependency in $K$ and the minimal amplitude of the effective hopping rate via the value of the asymmetry ratio $T_1/T_2$, where for the experiment $T=T_1+T_2=$\,1\,ms was chosen.

\subsubsection{TIME-OF-FLIGHT MEASUREMENTS:}

To reveal the quasi-momentum distribution of the BEC in the 1D lattice, the lattice and the dipole trap potential are suddenly turned off. We measure the absolute position of the interference pattern after 27\,ms time-of-flight, with the quasi-momentum distribution obtained without forcing as a reference.
When taking those pictures in the laboratory frame, one observes the quasimomentum distribution of the state evolving according to the time-independent effective Hamiltonian, enveloped by the momentum distribution of the lowest-band Wannier-function that oscillates like $m\dot x$. Fig. \ref{fig_SM} shows those oscillations over two periods of the forcing, corresponding to a Peierls phase of 0.64\,$\pi$.
\newline
For the data presented in the text, the pictures were always taken at times within a period where $\dot x =0$ and averaged over 3 to 7 experimental runs.

\end{document}